\documentclass[11pt,a4paper]{article}
\usepackage{hyperref}
\usepackage{times}
\usepackage{latexsym}

\usepackage{graphicx}
\usepackage{xcolor}
\usepackage[margin=1.1in]{geometry}

\title{Embedding-based Scientific Literature Discovery \\ in a Text Editor Application}

\author{Onur G\"{o}k\c{c}e, Jonathan Prada, Nikola I. Nikolov, Nianlong Gu, Richard H.R. Hahnloser\\
 Institute of Neuroinformatics, University of Zurich and ETH Zurich, Switzerland \\
  \texttt{\{onur, johny, niniko, nianlong, rich\}@ini.ethz.ch}
  }

\date{}

\begin{document}
\maketitle

\begin{abstract}
Each claim in a research paper requires all relevant prior knowledge to be discovered, assimilated, and appropriately cited. However, despite the availability of powerful search engines and sophisticated text editing software, discovering relevant papers and integrating the knowledge into a manuscript remain complex tasks associated with high cognitive load. To define comprehensive search queries requires strong motivation from authors, irrespective of their familiarity with the research field. Moreover, switching between independent applications for literature discovery, bibliography management, reading papers, and writing text burdens authors further and interrupts their creative process. Here, we present a web application that combines text editing and literature discovery in an interactive user interface. The application is equipped with a search engine that couples Boolean keyword filtering with nearest neighbor search over text embeddings, providing a discovery experience tuned to an author's manuscript and his interests. Our application aims to take a step towards more enjoyable and effortless academic writing.

The demo of the application\footnote{\url{https://SciEditorDemo2020.herokuapp.com/}} and a short video tutorial\footnote{\url{https://youtu.be/pkdVU60IcRc}} are available online.
\end{abstract}

\section{Introduction}

Writing is a complex problem-solving task that burdens authors with a high cognitive load \cite{Hayes-2012-ID1369}, which especially applies to inexperienced researchers \cite{Shah-2009-ID1370}. The typical workflow of composing an academic manuscript (be it a proposal, report, or paper) is an iterative process of conceptualizing ideas, formulating search queries, browsing search results, reading papers, eventually followed by assimilating and integrating the discovered knowledge. 

\begin{figure}[tb]
    \centering
    \includegraphics[width=0.5\columnwidth]{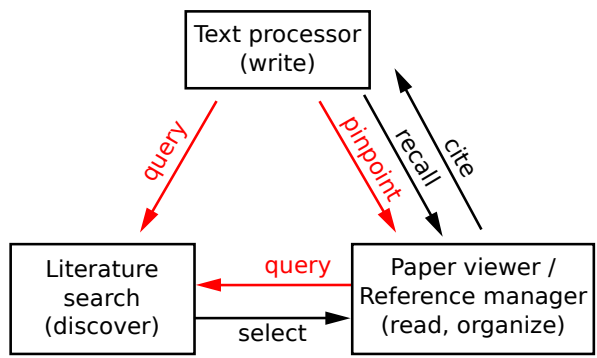}
    \caption{The typical workflow of scientific writing is largely based on independent software tools (text processor, reference manager, literature search engine, and paper viewer) that draw on diverse cognitive processes (recalling and citing articles, as well as searching, retrieving, and reading articles, black). Our web application focuses on assisting authors in literature discovery and in pinpointing relevant text passages in a paper (red).}
    \label{fig:sci_workflow}
\end{figure}

The current toolbox of scientific writing consists of text editors, search engines, reference managers, and paper viewers. These components are typically independent applications with limited interactivity. Consequently, authors are forced to navigate through diverse user interfaces repeatedly and need to link different parts of their workflow manually. We believe that there is a need for technology that makes literature discovery a seamless extension of the writing experience (Figure~\ref{fig:sci_workflow}).

Implicitly, each scientific statement requires an in-depth search for supporting or conflicting findings in the literature. Accordingly, authors must retain a strong motivation to iterate through many combinations of search terms even when the apparent gain from the search becomes sub-optimal \cite{Azzopardi-2018-ID1385}. In addition, the keywords intended for traditional search engines can be intrinsically biased because authors seek confirmation \cite{Nickerson-1998-ID1326} or because of gaps in their knowledge \cite{Athukorala-2013-ID1374}. The use of synonymous terminology, such as with the names of species in botany \cite{Rivera-2014-ID1373} or field-specific nomenclature \cite{Hodges-2008-ID1372}, further complicates formulating comprehensive search queries. Last but not least, the exponential increase in the number of scientific publications \cite{Larsen-2010-ID1371} makes it increasingly difficult to keep track of the literature and to incorporate new findings into one's work. 

Such challenges call for novel tools to alleviate the obstacles faced by authors. We, therefore, set out to design a workflow that simplifies the exploration of the scientific literature by making use of advances in natural language processing (NLP). We introduce a web application for writing scientific text with integrated literature discovery, paper reading, and bibliography management capabilities. 

Our application allows authors to retrieve papers that are similar to their manuscript (or to some of its parts) by utilizing text embeddings (Section \ref{sec:ranking}). In addition, the authors can confine the scope of retrieved papers to specific interests by applying keyword-based Boolean filters (Section \ref{sec:filtering}). Finally, to guide the authors in skim reading, similar sentences can be automatically highlighted in the retrieved papers. With these features, we aim to make the processes of literature discovery and scientific writing more efficient and enjoyable.

\section{Related Work}

\subsection{Platforms for Literature Search, Discovery, and Reference Management}

Currently, there are many independent applications for searching for and sharing of publications (e.g., Google Scholar, Pubmed, Web of Science, Meta, ResearchGate, and Iris.AI), for managing bibliography (e.g., Mendeley, Readcube, Paperpile, EndNote, and F1000), and for processing text (e.g., Microsoft Word, Google Docs, Overleaf, Dropbox Paper, and Sciflow). However, end-to-end applications that combine text editing with NLP-powered interactive literature discovery are scarce. Traditionally, text processors can interact with external software to search for content, to manage references, or to improve writing style via plug-ins, but such interactions are typically limited. 

A recent application, Raxter.io, provides a single interface for document writing and literature searching. Although Raxter.io allows fine-tuning of document-based search queries, its methods are not fully disclosed, and it neither supports flexible keyword definitions nor the automatic highlighting of relevant passages. Raxter.io also does not display the full body of papers unless the users manually import them.

\subsection{Methods for Literature Discovery}

Traditional search engines use a bag-of-words model with a frequency-based ranking function such as BM25 \cite{Robertson-2009-ID1382} to retrieve documents that match a query of one or more search terms. Obtaining useful search results requires well-formulated search queries \cite{Aula-2003-ID1436}, which can be a challenging task during exploratory search \cite{Belkin-2000-ID1435} and constitutes a cognitive load \cite{Gwizdka-2010-ID1437} that our application aims to ease.

Document similarity search methods \cite{Wan-2008-ID1381}, by contrast, use entire documents as the search queries, circumventing the need to define keywords for the search. State-of-the-art methods for retrieving similar documents rely on text embeddings \cite{Conneau-2018-ID1361,Adi-2016-ID1379,le2014distributed} and on efficient approximate nearest neighbor search algorithms \cite{Johnson-2017-ID1380}. However, embedding-based search methods seem rather inflexible in refining searches, because it is unclear how to steer search results in a particular direction without painstakingly having to modify the query document. 

Both keyword- and embedding-based search methods provide unique advantages, but there have not been many attempts at combining these methods to overcome their respective limitations.

\section{Literature Discovery}
 
\begin{figure}[tb]
    \centering
    \includegraphics[width=0.5\columnwidth]{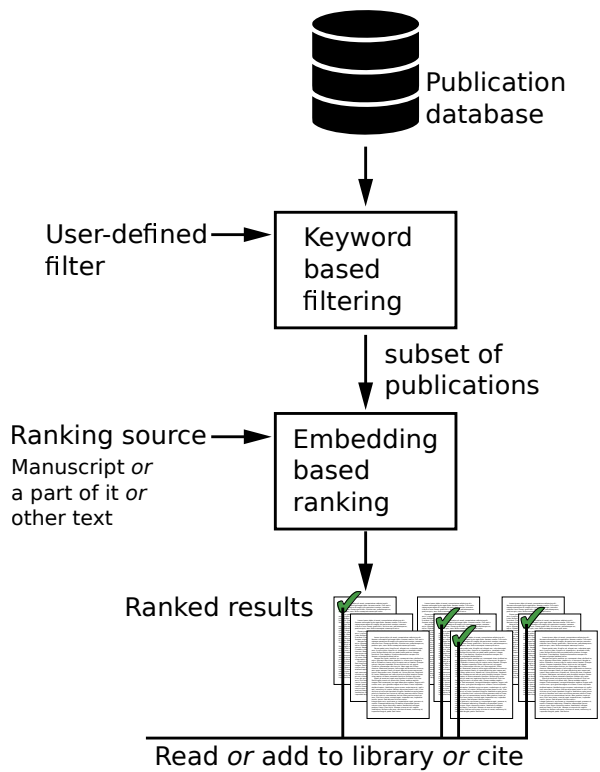}
    \caption{Overview of the literature discovery pipeline in our application. The search engine first filters our database for papers that match a set of user-defined keywords, and then ranks the filtered results according to their embedding-based proximity to a ranking source, such as an entire user manuscript. The top-ranked papers are presented to the user who can then save, cite, or read them, with the possibility of highlighting the most relevant sentences.}
    \label{fig:discovery}
\end{figure}

The pipeline for literature discovery in our application consists of two steps (Figure~\ref{fig:discovery}). First, the search engine retrieves a subset of the papers from our database that match a user-defined keyword-based filter. Second, the search engine ranks the filtered papers according to their similarity to the manuscript using document embeddings. We describe each of the two steps in detail below. Our database contains 2.7M papers from the Pubmed Central Open-Access subset (PMC-OA)\footnote{\url{https://www.ncbi.nlm.nih.gov/pmc/tools/openftlist/}}.

\subsection{Keyword-based Filtering} \label{sec:filtering}

An embedding-based search might return many papers that are similar to the manuscript but are of limited interest to the author. For example, authors of a medical manuscript on \emph{lung cancer} may seek similar treatments in the literature for another organ, but embedding-based ranking might retrieve papers only on lung cancer. The keyword-based filter can, in such cases, be used to restrict the ranking operation either to the papers mentioning that \emph{other organ} or to papers that do not mention \emph{lung}. Thus, filtering allows an author to focus the nearest neighbor search on the target keywords or on their absence. 

The filtering operation uses an inverted index of all unigrams in the database after the removal of stop words and word stemming (snowball) using the NLTK library\footnote{\url{https://www.nltk.org/}}. The resulting index has a dictionary size of 9.61M unigrams and requires $\sim$4 GB of memory.

\subsection{Embedding-based Ranking} \label{sec:ranking}

The ranking operation uses the document embeddings of the papers in our database. Given a \textit{ranking source} such as a paragraph or the entire manuscript, ``embedding-based ranking" sorts the papers returned by the keyword-based filter according to their cosine distance to the embedding of the ranking source. In other words, embedding-based ranking performs a brute-force nearest neighbor search on a subset of papers. The embedding of the ranking source is computed on demand whenever a search is performed.

As the document embedding model, we use Sent2Vec \cite{Pagliardini-2018-ID1335} because of its simplicity, speed, and good performance on various benchmark datasets \cite{Pagliardini-2018-ID1335,Nikolov-2019-ID1387}. The model has 400 dimensions and is trained on the PMC-OA corpus using a unigram vocabulary of $\sim$0.75M terms. After the training, we pre-compute the embeddings of all papers in our database and keep them in memory, which requires $\sim$4 GB. 

To test the performance of our model, we performed experiments on a simple text retrieval task. The goal of this task was to retrieve the full body of a parent paper given its abstract as the search query. We randomly sampled 10000 abstracts from the database and retrieved the 20 most similar papers for each abstract. As an evaluation metric, we counted the fraction of retrievals in which the parent paper appeared on top or among the top 20 results. Our model retrieved the correct parent paper as the top search result in 83.1\% of the trials, compared to 71.0\% when using a Sent2Vec model trained on Wikipedia \cite{Pagliardini-2018-ID1335}. Furthermore, the parent paper was among the top 20 retrievals in 95.1\% of cases when using our model, compared to 87.0\% for the Wikipedia Sent2Vec model. The higher retrieval performance of our model in this task likely arises from its training on a domain-specific corpus that contains rare words and terminologies \cite{Roy-2017-ID1391,blagec2019neural}. This suggests that the model would need to be retrained at regular intervals, particularly when papers from other domains are added to the database.

We have not systematically analyzed the retrieval performance when the query is formed by merely a part of the manuscript such as a block of a few sentences \cite{Gong-2018-ID1429,DeBoom-2015-ID1430}. We leave a detailed exploration of the effects of the query length on performance to future work.

\subsection{Scalability of Literature Discovery}

Although fast and efficient approximate nearest neighbor methods exist for retrieving the K nearest neighbors of a query vector, such schemes apply to ranking only, but not to the joint filtering and ranking steps (when nearest neighbors are sought among a subset of embeddings from the database). For this reason, in our search engine, there is no simple alternative to brute force search. Nevertheless, we find that retrieval is sufficiently fast, largely because the filtering step reduces the number of neighbors that need to be ranked. In future work, we will explore optimizations of the search engine, such as using approximate hashing techniques \cite{Datar-2004-ID1396,Norouzi-2012-ID1395}.

\section{User Interface and Workflow}

\begin{figure*}[tb!]
    \centering
    \includegraphics[width=0.7\textwidth]{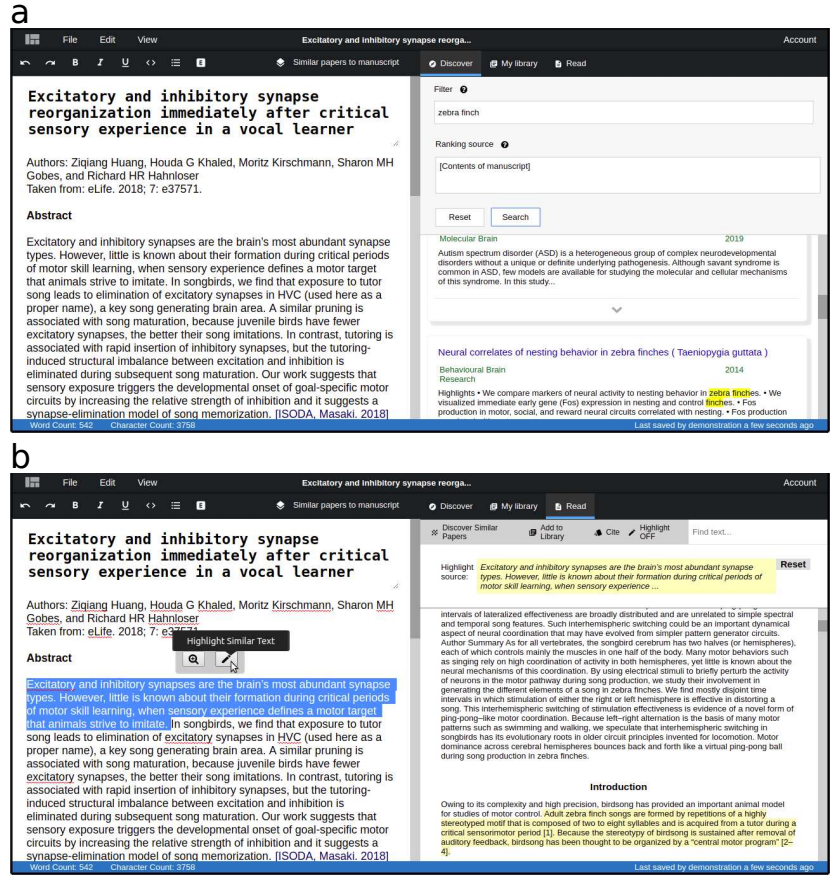}
    \caption{The user interface of the application. a) the \textit{Discover} tab lists the retrieved papers that are similar to the manuscript. b) the \textit{Read} tab allows users to view papers and to highlight the sentences that are similar to the selected text in the manuscript.}
    \label{fig:frontend}
\end{figure*}

The user interface (UI) consists of \textbf{(1)} a \textit{text editor} that provides basic functionality for drafting a manuscript, such as loading \& saving documents, formatting text, and inserting \LaTeX{} equations, code snippets, or bullet points (Figure \ref{fig:frontend}a, left), and \textbf{(2)} a \textit{literature explorer} encompassing multiple components, which can be accessed on their respective tabs (Figure \ref{fig:frontend}a, right):

\begin{itemize}
    \item\textbf{\textit{Discover}} for performing searches and browsing the search results to discover relevant literature
    \item\textbf{\textit{My Library}} for managing the user bibliography and for citing papers in the manuscript
    \item\textbf{\textit{Read}} for paper viewing and for actions that facilitate literature exploration, such as discovering similar papers to the one being viewed and highlighting the sentences in the paper that are similar to the selected text in the manuscript (Figure \ref{fig:frontend}b, right)
\end{itemize}

A search can be initiated without keyword filters by clicking the ``Similar papers to the manuscript" button located above the text editor. As a result, the 1000 most similar papers are listed in the \textit{Discover} tab with their metadata (title, authors, journal, publication year, and abstract). 

A more granular search can be performed by selecting a section (e.g., sentences, paragraphs) from the manuscript, which reveals a hovering menu over the selected text (visible in Figure~\ref{fig:frontend}b). Clicking on the magnifying glass icon on this menu performs a search using the selected text as the ranking source and consequently returns the papers similar to the selected text.

To steer discovery towards a particular set of terms, the user can define a keyword-based Boolean filter using the format \verb"term1 term2|term3 !term4" to confine the results to those papers that contain \emph{term1} and (\emph{term2} or \emph{term3}), but not \emph{term4}. 

Clicking on a search result displays the content of the paper in the \textit{Read} tab. In this tab, the user finds additional actions above the viewed paper to interact with it.

If, after viewing the paper, the user finds it interesting, then pressing the ``Add to Library" button saves the paper in the user bibliography, which can be viewed under the \textit{My Library} tab. Alternatively, the ``Cite" button places a reference to the paper at the current cursor position in the text editor and adds the paper to the user bibliography. Inserted references in the manuscript are links, and clicking on them conveniently opens the respective paper in the \textit{Read} tab. Deleting the link removes the reference from the manuscript.

To facilitate the exploration of the literature further, the \textit{Read} tab contains additional functions: ``Discover similar papers" performs a search using the viewed paper as the ranking source. If a filter is already present in the \textit{Discover} tab, then the search results are filtered accordingly. The ``Highlight" button highlights the 20 sentences in the viewed paper that are most similar to the ranking source, i.e., similar to the query of the last search performed on the application. Alternatively, the user can select a part of the manuscript and press the marker icon on the revealed hovering menu (Figure~\ref{fig:frontend}b) to highlight the sentences that are most similar to the selection. The highlighting feature computes the embedding of each sentence in the viewed paper to assess similarity. The "Find Text" field uses the web-browser's built-in \textit{find} functionality to match the value of the field with the viewed paper.

The \textit{My Library} tab lists all the papers in the user bibliography. Ticking the ``Cited content only" box filters this list to show only the papers cited in the manuscript. The user can press the ``Cite" button next to a paper to insert a reference to the paper at the cursor position in the text editor. The user can also add papers to the library manually by entering the digital object identifier of the paper in the form that appears upon pressing the ``Manual entry" button. Items under \textit{My Library} can be removed by clicking on the ``Remove" button next to the item.

\section{Conclusion}

We have described an application that aims to reduce the manual workload involved in exploring the scientific literature. Our application combines the processes of reading papers and of writing scientific manuscripts into a single user interface and links them using NLP algorithms. 

In future work, we will focus on expanding the database to include additional domains and article sources. We will work on augmenting the workflow with automated tasks, such as suggesting references as the author writes a manuscript, or notifying users about the latest publications relevant to their work. We will also seek to improve discovery performance by testing more recent text embedding methods (e.g., BERT \cite{Devlin-2018-ID1345}) and by optimizing the search for different input text lengths, such as a whole document, a paragraph, or even a single sentence. 

Finally, we are aware that keyword-based Boolean filtering might be prone to the same biases and challenges inherent in the traditional search queries, as discussed above. We will investigate whether query expansion techniques \cite{Azad-2019-ID1428} could mitigate this issue by suggesting or automatically appending semantically related keywords to the Boolean filters.

\section*{Acknowledgements}

We acknowledge support from the Swiss National Science Foundation (grant 31003A\_156976). We also thank the anonymous reviewers for their useful comments. 

\bibliographystyle{plain}
\bibliography{main}

\end{document}